\documentclass[prd,twocolumn,groupedaddress,nofootinbib]{revtex4}

\usepackage{graphicx}

\def\dim{\hbox{dim}\,}
\def\vev#1{\left\langle #1 \right\rangle}
\def \cov {\bigtriangledown}
\def\bfx{{\mathbf r}}

\def\OMIT#1{}

\begin{document}
\bibliographystyle{apsrev}

\preprint{\vbox{\hbox{HUTP-01/A052} 
                \hbox{UCSD/PTH 01-18}  }}

\title{Spontaneously Broken Spacetime Symmetries and Goldstone's Theorem}


\author{Ian Low$^a$ and Aneesh V. Manohar$^b$}
\affiliation{$^a$Jefferson Physical Laboratory, Harvard University, 
Cambridge, MA 02138\\
$^b$Department of Physics, University of California at San Diego,
La Jolla, CA 92093}
%

\date{30 October 2001}

\begin{abstract}
Goldstone's theorem states that there is a massless mode for each broken
symmetry generator.  It has been known for a long time that the 
naive generalization of this counting fails to give the correct number of
massless modes for spontaneously broken spacetime symmetries. We explain how to
get the right count of massless modes in the general case, and discuss examples
involving spontaneously broken Poincar\'e  and conformal invariance.
\end{abstract}

\pacs{}

\maketitle


The proof of Goldstone's theorem for internal symmetries is now standard
material in many textbooks on quantum field theory. Briefly stated, the theorem
asserts that for a physical system with a global internal symmetry group $G$
which is spontaneously broken down to a subgroup $H$, there is a massless mode
corresponding to each broken generator~\cite{goldstone}. In other words, the
number of Goldstone bosons is equal to the dimension 
$\dim(G/H)=\dim(G)-\dim(H)$ of the coset space $G/H$.\footnote{For the purposes of this paper, 
we will ignore subtleties such as the Coleman-Mermin-Wagner theorem on Goldstone bosons in two dimensions.} Moreover, if the global
symmetry $G$ is gauged into a local symmetry group, some of the gauge bosons
become massive through the Higgs mechanism. In this case, the number of massive
bosons, which equals the number of  would-be Goldstone bosons, is still
$\dim(G/H)$.

The naive generalization of this counting fails to give the correct number of
massless modes for spontaneously broken spacetime symmetries.  There are at
least two well-known cases.  One is the spontaneous breaking of rotational and
translational invariance due to, for example, an extended object such as a
domain wall or $D$-brane. In examples discussed in
Ref.~\cite{Polchinski:1992ed,Sundrum:1999sj}, there are massless modes
corresponding to only the broken translational generators. The second example
is the spontaneous breaking of conformal symmetry down to Poincar\'e symmetry.
In four dimensions, the conformal group has 15 generators, whereas the
Poincar\'e group has 10 generators. Naive counting using $\dim(G)-\dim(H)$
would give 5 Goldstone modes. However, as was discussed in
Ref.~\cite{Isham:1970gz,volkov}, there is actually only one massless mode,
corresponding to the dilatation generator.

It is known that in non-Lorentz-invariant theories, the number of massless
modes can be less than the number of broken generators even for internal
symmetries, and the rule for counting the massless modes is given in
Ref.~\cite{Nielsen:1976hm}. We analyze a different problem---that of counting
massless modes in Lorentz-invariant theories with broken spacetime symmetries.
Following a review of the two well-known examples of spontaneously broken
Poincar\'e and conformal symmetries, we present the criterion for counting
massless modes, and show how the same result can be derived by applying the
coset construction of spontaneously broken
symmetries\cite{volkov,Coleman:1969sm}.



A large class of models with extra dimensions consider quantum fields confined
in a $(p+1)$ dimensional hypersurface,  which is generally called a $p$-brane,
embedded in a $d$ dimensional spacetime, with $d>p$.  If the vacuum state of a
theory in $d$-dimensional flat space contains  a $p$-brane,  the $d$
dimensional Poincar\'e group, denoted by $ISO(d-1,1)$\footnote{We use the
Minkowski metric  $\eta_{MN}=(-,+,+,\cdots)$.}, is spontaneously broken down to
the $(p+1)$  dimensional Poincar\'e group $ISO(p,1)$.  We will consider this
pattern of symmetry breaking; the spontaneous breaking of Poincar\'e invariance
by vortices or domain walls are special cases.

The indices of the bulk spacetime will be denoted by capital Roman letters,
$M,N=0,\cdots, d-1$, the $p$-brane coordinate indices by Greek letters,
$\mu,\nu=0,\cdots,p$, and the remaining $d-(p+1)$ of the bulk indices, by lower
case Roman letters $m,n=p+1,\cdots,d-1$. The coordinates of the bulk spacetime
are $X^M$, and those intrinsic to the $p$-brane are $x^\mu$. The translation
generators $P^M$ can be divided into two sets, $P^\mu$ which remain unbroken,
and $P^m$, which are broken by the $p$-brane. The Lorentz generators  $J^{MN}$
split into the unbroken generators $J^{\mu\nu}, J^{mn}$ and the broken
generators $J^{\mu n}$. The position of a point $x$ on the $p$-brane is
described by the bulk coordinates $Y^M(x)$. It is possible to choose a gauge
such that $Y^\mu(x)=x^\mu$ and the remaining components, $Y^m(x)$, can be
thought of  as the Goldstone modes  corresponding to the broken translational
generators\cite{Sundrum:1999sj}, which describe the fluctuations of the
$p$-brane in the transverse directions.  The number of Goldstone modes is
$d-p-1$, which is the same as the number of broken translation generators
$P^m$. There are no additional Goldstone modes corresponding to the broken
Lorentz generators $J^{\mu n}$.


Next consider the spontaneous breaking of conformal
symmetry\cite{Isham:1970gz}. Any  Lagrangian with a symmetry group $H$ can be
made $G$ invariant, $G\supset H$, by adding Goldstone bosons so that it appears
the symmetry $G$ is spontaneously broken. When $G$ is taken to be the conformal
group and $H$ the Poincar\'e group, although the broken generators are the
dilatation and special conformal transformations, we only need one massless
mode $\sigma(x)$, the dilaton, to make the Lagrangian conformally
invariant.\footnote{This is related to the fact that a theory which is scale
invariant is also conformal invariant if a certain condition is satisfied,
which is true for a wide class of
theories\cite{Callan:1970ze,Polchinski:1988dy}.}

As a simple example, consider a scalar  $\phi^4$ theory in four dimensions 
which can be made conformally invariant by adding the dilaton $\sigma(x)$  in
the following way:
\begin{eqnarray}
S&=& \int d^4x \Bigl[\frac12 (\partial_\mu+f\, \partial_\mu\sigma)\phi
 (\partial^\mu+f\, \partial^\mu\sigma)\phi
+\Lambda e^{-4f\sigma}\nonumber\\ 
&&-\frac 12 m^2 \phi^2 e^{-2f\sigma}-
 \frac{\lambda}4 \phi^4 + \frac12 e^{-2f\sigma}
\partial_\mu\sigma \, \partial^\mu\sigma \Bigr].\label{conf}
\end{eqnarray}
Note that, under a scale transformation $x
\to e^{-d} x$, the field $\sigma$ transforms in a non-linear way $\sigma(x) \to
\sigma(e^{-d} x) - d/f$ and is indeed the Goldstone mode corresponding to the
dilatation. The Lagrangian~Eq.~(\ref{conf}) describes a theory with
spontaneously broken conformal symmetry, with one massless mode coupling to the
dilatation current. There are no additional massless modes corresponding to the
breaking of the special conformal transformations.

Assume that a symmetry group $G$ with $\dim G$ generators $T^A$
(capital Roman superscript) is  broken down to a symmetry group $H$ with $\dim H$
generators $T^\alpha$ (Greek superscript). The remaining $\dim G - \dim H$
generators $T^a$ (lower case Roman superscript) are referred to as the broken
generators. Let $\phi(\bfx)$  be the symmetry breaking order
parameter, $T^\alpha \vev{\phi(\bfx)}=0$, and  $T^a  \vev{\phi(\bfx)}\not=0$. 
In the case of internal symmetry breaking $\phi(\bfx)$ is a scalar
field, but for spacetime symmetry breaking, $\phi(\bfx)$ can be a tensor field.


Consider first the case of a broken internal symmetry. The massless modes
are small amplitude long-wavelength fluctuations of the order parameter,
\begin{equation}\label{5}
\delta \phi(\bfx) =  c_A(\bfx) T^A \vev{\phi (\bfx)}
=  c_a(\bfx) T^a \vev{\phi (\bfx)},
\end{equation}
where $c_a(\bfx)$ is now a slowly varying function of $\bfx$.
The generators $T^\alpha$ corresponding to the unbroken generators do not
generate massless excitations, since $T^\alpha \vev{\phi(\bfx)}=0$. The
remaining $c_a$ can be chosen independently, and the number of independent
modes is clearly the same as the number of broken generators, $\dim (G/H)$.

For spontaneously broken spacetime symmetries, the number
of massless modes is no longer equal to the number of broken generators. 
Massless modes are still given by small
amplitude long-wavelength fluctuations of the order parameter, Eq.~(\ref{5}), where $c_a$ can depend on the coordinates $\bfx$ in the directions in which translation remains unbroken.
The number of independent massless modes is the number of broken
generators $\dim(G/H)$ minus $n_x$, the number of independent solutions to 
\begin{equation}\label{spacetime} 
c_a(\bfx) T^a \vev{\phi (\bfx)} =0.
\end{equation} 
The key point is that $n_x \ge 0$: there can be non-trivial solutions to
Eq.~(\ref{spacetime}) when $c_a$ and $T^a$ both depend on $\bfx$.
The generators $T^a$ are linearly independent, but the long-wavelength fluctuations they produce
need not be. The number of Goldstone bosons is then
$\dim(G/H)-n_x$, and is reduced from the naive counting of broken generators.
Equation~(\ref{spacetime}) can always be used to determine $n_x$, even in the
case of internal symmetries. If the generators $T^a$ are internal generators,
Eq.~(\ref{spacetime}) has no non-trivial solutions, and $n_x=0$.

It is easy to see how there could be non-trivial solutions to Eq.~(\ref{spacetime}), 
thus reducing the number of
Goldstone modes. Consider in three dimensions, a ground state with an
infinitely long, straight string parallel to the $y$ axis, as shown in  Fig.~1.
The three-dimensional Poincar\'e group is spontaneously broken to the 
two-dimensional Poincar\'e group. A rotation in the $x$-$y$ plane changes the
orientation of the string, whereas a translation in the $x$ direction shifts
the string parallel to itself. The effect of these two symmetry operations are
apparently very different. Nevertheless, if we perform a {\em local}
translation on the string, in the sense that the amount of translation is
different at every point on the string, the effect is to produce a bump on the
string. This bump can clearly be compensated by performing a {\em local}
rotation on the string (see Fig.~1).

\begin{figure}[t]
\includegraphics[width=8cm]{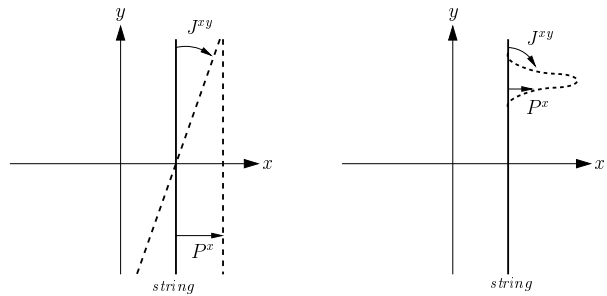}
\caption{A ground state with a string  breaks the three-dimensional
Poincar\'e group down to the two-dimensional Poincar\'e group. Global translation
and rotation on the string are distinctly different, whereas the effects of
local translation and rotation on the string can be made the same.}
\end{figure}

The translational symmetry breaking of $P_x$ produces
massless modes,
\begin{equation}
\delta \phi(\bfx)=\epsilon(y) P_x \vev{\phi(\bfx)},
\end{equation}
where $\epsilon$ only depends on $y$, the coordinate in the direction of the
unbroken translation generator $P_y$.\footnote{Goldstone modes can only
propagate in the direction of the unbroken translations. They have a dispersion
relation $\omega(k)$ with $\omega(k)\to 0$ as $k\to 0$. $k$ is only defined in
the translationally invariant directions. The broken translation $P_x$
generates translational zero-mode describing the fluctuations of the string
in the $x$ direction.} Similarly, rotational symmetry breaking gives the mode
\begin{equation}
\delta \phi(\bfx)=\theta(y) J_{xy} \vev{\phi(\bfx)},
\end{equation}
where $\theta$ only depends on $y$. Requiring these two operations to exactly
cancel each other, we have
\begin{eqnarray}
 \epsilon(y) P_x \vev{\phi(\bfx)}  &=& 
-y\theta(y) P_x\vev{\phi(\bfx)} \label{2dexample}
\end{eqnarray}
where we have used $P_y \vev{\phi(\bfx)} =0$ and the relation
 $J_{xy}= x P_y-y P_x$ valid for spinless
particles.\footnote{More precisely, one uses the relation $M^{\mu\nu\lambda}(x)
=x^\nu T^{\mu \lambda}(x)- x^\lambda T^{\mu \nu}(x)$ between the stress-tensor
and the angular momentum density.} Equation~(\ref{2dexample}) is clearly
satisfied by choosing $\epsilon(y)=-y\theta(y)$. Note that no solution would be
possible if $\epsilon$ and $\theta$ were both chosen to be constants. In this
example $P_x$ and $J_{xy}$ do not generate independent massless excitations,
and $n_x > 0$.

In the general case, acting on Eq.~(\ref{spacetime}) with the unbroken
translation $P_\mu$ gives
\begin{eqnarray}
 0&=& P_\mu\, c_a(\bfx) T^a \vev{\phi(\bfx)}=
 \left[P_\mu, c_a(\bfx) T^a \right]\vev{\phi(\bfx)}
 \nonumber \\
&=& -i\left(\partial_\mu c_a(\bfx) T^a -  f^{\mu a b}c_a(\bfx) T^b\right)
\vev{\phi(\bfx)},
\label{pde}
\end{eqnarray}
where we have written the commutator in the most general form
\begin{equation}
\label{commutator}
\left[ P_\mu, T^a\right] = i f^{\mu ab}\ T^b + i f^{\mu a \beta }\ T^\beta.
\end{equation}
The $T^\beta$ are unbroken generators and thus annihilate the vacuum. If the
$T^a$ are internal generators, $\left[P_\mu, T^a\right]=0$, and
Eq.~(\ref{pde}) implies that $c_a$ are constant, so that Eq.~(\ref{spacetime})
has no non-trivial solutions.

As long as there are some non-zero $f^{\mu ab}$, the non-trivial solution
satisfies
\begin{equation}
\label{count}
\left(\partial_\mu c_a(\bfx) -  c_b(\bfx) f^{\mu ba}\right)T^a\vev{\phi(\bfx)}  = 0.
\end{equation}
This is equivalent to saying that the Goldstone mode for $T^b$ and the
gradient of the Goldstone mode for $T^a$ are linearly dependent, so they do not generate independent massless excitations.
The non-trivial solutions to Eq.~(\ref{count})
reduce the number of Goldstone bosons, and there is a one-to-one correspondence
between the non-trivial solutions of Eq.~(\ref{count}) and
Eq.~(\ref{spacetime}). We will see that Eq.~(\ref{count}) also occurs in the
coset construction of the low energy effective theory.

Propagating Goldstone modes exist when there are unbroken translational directions. It is also interesting to consider configurations which break all the translational
invariance; for example a soliton such as a magnetic monopole or Skyrmion. In this case, one has zero modes that correspond to changes in the collective coordinates of the soliton.
The counting here is the same as in the case for internal symmetry, and there
is no relation between the translational and rotational generators. We can see
this from Eq.~(\ref{spacetime}) as well. We stressed that the spacetime
dependence of the coefficients $c_a(\bfx)$ is on the coordinate in the unbroken
translation. When all the translations are broken, $c_a(\bfx)$ have no
spacetime dependence and are purely constants, Eq.~(\ref{spacetime}) has no non-trivial solutions, and the counting is the same as for internal symmetries. 

To see how the counting of Goldstone modes works for the two examples discussed
earlier, let us write down the full conformal algebra:
\begin{eqnarray}
\left[J_{MN},J_{PQ}\right]&=&i(\eta_{NP}J_{MQ}-\eta_{MP}J_{NQ}\nonumber\\
&&\qquad -\eta_{NQ}J_{MP}
           +\eta_{MQ}J_{NP}) \\
\left[J_{MN},P_Q\right]&=&i(\eta_{NQ}P_M-\eta_{MQ}P_N) \label{trans}\\
\left[J_{MN},K_Q\right]&=&i(\eta_{NQ}K_M-\eta_{MQ}K_N) \\
\left[P_M,K_N\right]&=& 2i J_{MN} -2i\eta_{MN}D  \label{specialcon}\\
\left[D,K_M\right] &=& i K_M \\
\left[D,P_M\right] &=& -i P_M,
\end{eqnarray}
where $D$ is the generator for dilatation, and $K^M$ are the generators for
special conformal transformations. If the $d$ dimensional Poincar\'e group is
broken down to the $(p+1)$ dimensional Poincar\'e group due to the presence of
the $p$-brane, we have from Eq.~(\ref{trans})
\begin{equation}
\left[P_\mu, J_{\nu m}\right]\ \vev{\phi(\bfx)} = i\eta_{\mu\nu} P_m\
\vev{\phi(\bfx)}, 
\end{equation}
where $J_{\mu m}$ and $P_m$ are the broken rotational and translational
generators, respectively. Therefore  $J_{\nu m}$ and $P_m$ do not generate independent Goldstone modes.
Similarly, for the
conformal group spontaneously  broken down to the Poincar\'e group, we have
from Eq.~(\ref{specialcon})
\begin{equation}
\left[P_M,K_N\right] \ \vev{\phi(\bfx)} = -2i\eta_{MN}D  \ \vev{\phi(\bfx)}
\end{equation}
and therefore all the modes for $K_N$ can be eliminated, leaving only the
dilaton. 


Next we apply the coset construction for theories with spontaneous
symmetry breaking, introduced in Ref.~\cite{Coleman:1969sm} for internal
symmetries and modified in Ref.~\cite{volkov} for spacetime symmetries, to the
discussion of counting massless modes. It is convenient to divide the unbroken generators $T^\alpha$ into the unbroken  momenta $P_\mu$, and the rest, $V_s$. Consider the group element
\begin{equation}
\Omega(x,\xi)=e^{i x^\mu P_\mu} e^{i\xi^a(x) T_a},
\end{equation}
which transforms under the action of an element $g$ of $G$ as
\begin{equation}
g\ e^{i x^\mu P_\mu}\, e^{i\xi^a T_a} = e^{i x^{\prime \mu} P_\mu}\,
e^{i\xi^{\prime a}(x^\prime) T_a}\,  h(\xi^a(x),g), 
\end{equation}
where $h(\xi^a(x),g)$ is an element of $H$ depending on $\xi^a(x)$ and $g$. If
$g$ belongs to the unbroken group $H$, the transformation of $x^\mu$ and
$\xi^a(x)$ becomes linear. For example, if $g$ is one of the unbroken Lorentz
generators, it simply induces the usual Lorentz transformation $x^\prime
=\Lambda x$ and $\xi^{\prime}(x^\prime)= S^{-1}(\Lambda)\xi(x)$. However, under
a translation $e^{iy^\mu P_\mu}$, the spacetime coordinates always transform
inhomogeneously,\footnote{Recall that we can think of the spacetime coordinates
$x^M$ as parameterizing the coset (Poincar\'e)/(Lorentz).} $x^\prime=x+y$,
whereas $\xi^\prime(x^\prime)=\xi(x)$. This is why $P_\mu$ play the same role
as other broken generators in $\Omega(x,\xi)$. 

In order to construct an effective action invariant under the full symmetry
$G$, we need to consider the Maurer-Cartan one form
\begin{eqnarray}
\label{oneform}
\Omega^{-1}(x,\xi)\, d\Omega(x,\xi)
= i\left(\omega_P^\mu P_\mu
+\omega_T^a T_a +\omega_V^s V_s \right).
\end{eqnarray} 
The one forms $\omega_P^\mu$ and $\omega_T^a$ transform covariantly and are
related to the spacetime vielbeins and the covariant derivatives of the
Goldstone field $\xi_a$:
\begin{eqnarray}
\omega_P^\alpha &=& dx^\mu\, e^\alpha_\mu \label{vielbein}\\
\omega_T^a &=&  dx^\mu\, e_\mu ^\alpha\, \cov_\alpha \xi^a \label{goldstone}.
\end{eqnarray}
On the other hand, $\omega_V$ is the gauge field (sometimes called the spin connection)
associated with the unbroken group $H$,
\begin{equation}
\omega_V^s = dx^\mu\ \omega_{V\ \mu}^{\phantom{V}s} \label{gaugefield},
\end{equation}
and has the same transformation law as the gauge field under local
transformations of $H$.

In order to compute Eq.~(\ref{oneform}), we need the following commutation
relations, written in the most general form,
\begin{eqnarray}
\left[ P_\mu, T_a \right] &=& i f^{\mu a\nu}P_\nu + i f^{\mu ab}T_b + i
f^{\mu as}V_s , \label{general}\\
\left[ T_a, T_b \right] &=& i f^{ab\mu}P_\mu + i f^{abc}T_c +i f^{abs}V_s .
\end{eqnarray} 
Therefore $f^{\mu a\nu}$ and $f^{ab\mu}$ contribute to the spacetime
vielbeins Eq.~(\ref{vielbein}), $f^{\mu ab}$ and $f^{abc}$ contribute to
to the covariant derivative of the Goldstone boson Eq.~(\ref{goldstone}),
and $f^{\mu as}$ and $f^{abs}$ contribute to the spin connection
Eq.~(\ref{gaugefield}). Focusing on the Goldstone field, and working at
linearized order, we have
\begin{equation}
\label{linearized1}
\omega^a_T = ( \partial_\mu \xi^a -f^{\mu ba}\xi^b)dx^\mu.
\end{equation}
The effective Lagrangian contains $\Omega^{-1} d\Omega$ acting on $\vev{\phi}$,
so that the Goldstone boson fields occur via 
\begin{equation}
\label{linearized}
\omega^a_T T^a \vev{\phi}=( \partial_\mu \xi^a -f^{\mu ba}\xi^b)dx^\mu T^a
\vev{\phi}.
\end{equation}
Here we see the possibility of expressing some of the Goldstone modes in terms
of derivative of other Goldstone modes by setting Eq.~(\ref{linearized}) to
zero, which reduces the number of independent Goldstone modes that occur in the
effective Lagrangian. Note that the linearized covariant derivative is exactly
Eq.~(\ref{count}), the condition for non-trivial solutions to
Eq.~(\ref{spacetime}).

For the case of a $p$-brane breaking the Poincar\'e group spontaneously, we
chose to write $\Omega$ as
\begin{equation}
\label{poincare}
\Omega = e^{ix^\mu P_\mu}\ e^{iY^a(x) P_a}\ e^{i\theta^{\nu b}(x) J_{\nu b}}.
\end{equation}
The covariant derivative of the Goldstone mode $Y^a(x)$ is
\begin{equation}
\label{poincaremcform}
\omega_P^b =  \left(R(\theta)^b_\mu + R(\theta)^b_a \partial_\mu Y^a\right)
  dx^\mu,
\end{equation}
where
\begin{eqnarray}
\left(R(\theta)\right)_{MN} &=& \left( e^{i\theta^{\nu b}\Sigma_{\nu b}}
\right)_{MN} ,\\
\left(\Sigma^{\nu b}\right)_{MN}&=& 
i \left(\delta^\nu_M\ \delta^b_N - \delta^\nu_N\ \delta^b_M\right).
\end{eqnarray}
The covariant derivative of the Goldstone field $Y^a(x)$ involves $Y^a$ as well
as $\theta^{\nu b}(x)$, the Goldstone field for the broken rotational
generators. It is therefore possible to solve for $\theta^{\nu b}(x)$ in terms
of the derivatives of $Y^a(x)$ by setting the covariant derivative to zero. 

For the spontaneous breaking of conformal symmetry, we follow
Ref.~\cite{volkov} and write
\begin{equation}
\Omega = e^{ix\cdot P}\, e^{i\varphi(x)\cdot K}\, e^{i\sigma(x) D}.
\end{equation}
The covariant derivative of the dilaton is
\begin{equation}
\omega_D = (\partial_\mu\sigma+2\varphi_\mu)dx^\mu 
\end{equation}
Again we can replace the field $\varphi_\mu$ everywhere by $-(1/2)\partial_\mu
\sigma$ by setting the covariant derivative of the dilaton field to zero.  The
fact that one can eliminate some Goldstone fields this way is called the
inverse Higgs effect in Ref.~\cite{Ivanov:1975zq}. 

As a final note, it should be clear that choosing a different parameterization
of the coset space $G/H$ would give a different relation among the various
Goldstone modes. Nevertheless, the number of massless modes is determined by
the non-vanishing $f^{\mu ab}$ in Eq.~(\ref{general}), and the number of
Goldstone modes is independent of the parameterization of the coset.

I.L.~is supported in part by the National Science Foundation under grant number
PHY-9802709. He also acknowledges useful conversations with Nima Arkani-Hamed,
Andy Cohen, Tom Mehen, and John Terning. A.M.~is supported in part by the
Department of Energy under grant DOE-FG03-97ER40546. We are grateful to
Krishna Rajagopal for bringing Ref.~\cite{Nielsen:1976hm} to our attention.


\end{document}